# Classification of earthquakes main shocks

A.V. Guglielmi, O.D. Zotov

*Schmidt Institute of Physics of the Earth, Russian Academy of Sciences; Bol'shaya Gruzinskaya str., 10, bld. 1, Moscow, 123242 Russia; guglielmi@mail.ru (A.V.), ozotov@inbox.ru (O.D.)*

*Abstract*: The concept of a triad of tectonic earthquakes as a natural trinity of foreshocks, main shock and aftershocks is introduced. The basis for classifying the main shocks is the belonging of the main shock to one or another category of triads. The basis for classifying triads is the ratio between the number of foreshocks and the number of aftershocks that occurred during equal periods of time before and after the main shock. The existence of three classes of triads has been discovered, called classical, mirror and symmetrical triads. Each class is divided into two species, called the full and shortened triad. Thus, the existence of three classes and six species of main shocks is established. The classification is natural, is made on one basis and is proportionate, i.e. the sum of classes and the sum of species are equal to the volume of the generic concept of the triad. The discovery of a small but obviously non-empty set of mirror triads is an interesting and important concrete result of classification. Confidence is expressed that the classification of main shocks opens up new perspectives for earthquake research.

*Key words*: foreshocks, aftershocks, triad of earthquakes, taxonomy of triads, mirror triad.

# 1. Introduction

Classification of natural phenomena is widely used in natural science.

Classification allows us to express the diversity of phenomena in a limited number of well-identifiable species that differ from each other according to a certain characteristic. Earthquakes are classified primarily by magnitude, distinguishing between strong, weak and catastrophic earthquakes [Bolt, 1978]. An important indicator is also the depth of the hypocenter, which indicates whether the earthquake occurred in the crust or in the mantle. There are also continental and oceanic earthquakes, tectonic and volcanic, and so on. Systematization of this kind, which provides information about an event in a condensed form, is useful in both applied and fundamental seismology.

In this paper we propose to expand the systematics of the main impacts of tectonic earthquakes. We use a new feature of dividing earthquakes into three classes and six types. Let's explain what was said.

In the review [Guglielmi, 2015] the term *classical triad* was introduced to denote the natural trinity of foreshocks, mainshock and aftershocks. Observations show that before and after the main shock, for the same period of time, the number of foreshocks is much less than the number of aftershocks. Quite often, foreshocks are absent altogether. It is proposed to call such a triad *shortened*. An obvious question arose as to whether the tectonophysical conditions for the preparation of the main shock included in the full and shortened triads differ radically.

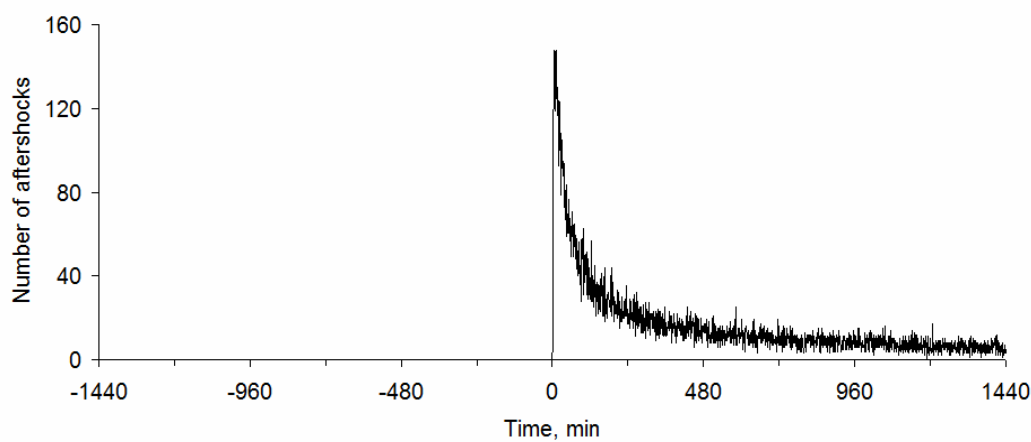

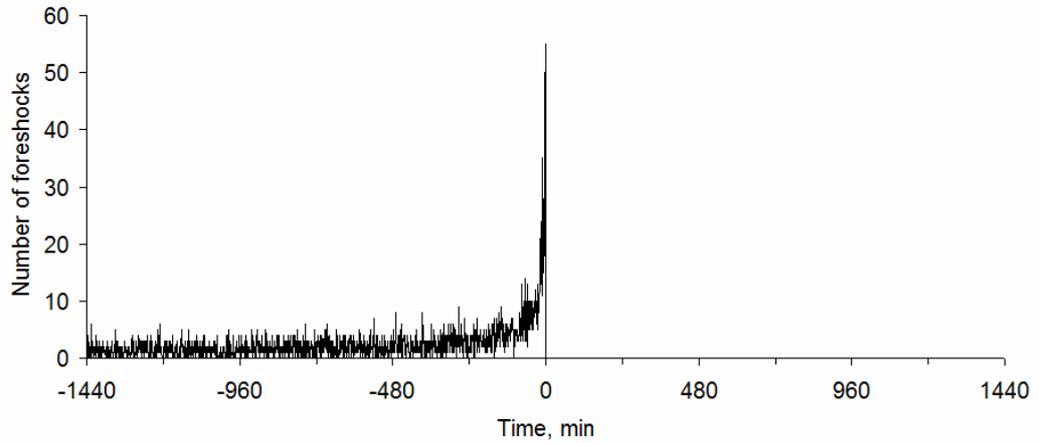

**Fig. 1**. Generalized classical and mirror triads. The graphs were obtained by synchronous summation of shortened triads for the period from 1973 to 2019 years with magnitudes of main shocks $5 \leq M < 7$. The depth of the hypocenters is less than 250 km. The moments of main shocks are marked by a vertical line.

The issue became more acute after *mirror triads* were discovered during the research process [Zotov, Guglielmi, 2021]. Figure 1 shows a generalized mirror triad (bottom) and a generalized classical triad (top). The figure was constructed using USGS/NEIC catalog data for 1973-2019. The main shocks with magnitudes in the range of $5 \leq M < 7$ and hypocenter depths not exceeding 250 km were selected.

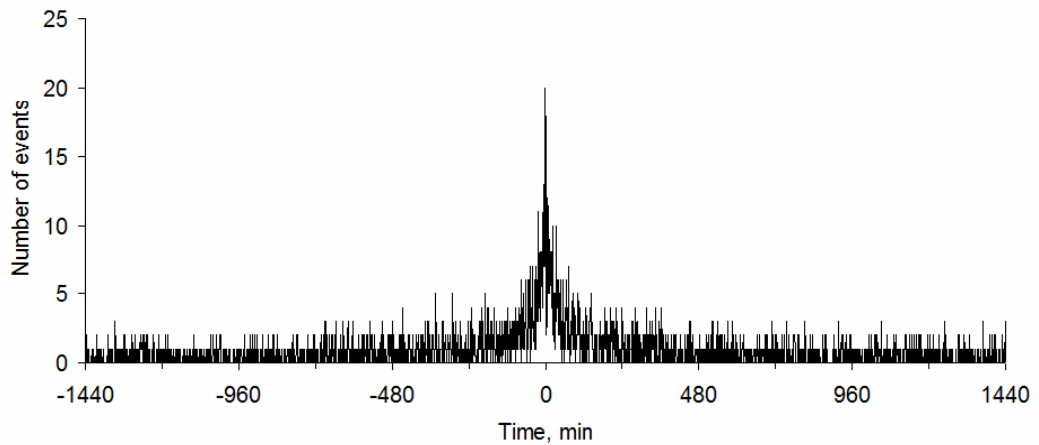

**Fig. 2**. Generalized complete symmetric triad for $5 \leq M < 7$.

To complete the picture, we performed a successful search for *symmetrical triads* in which the number of foreshocks is equal to the number of aftershocks (Figure 2). Thus, formally and in fact, there are three classes and six species of triads. We propose to classify the main attacks according to which of the specified triad categories a given main attack belongs to.

We will indicate the exact criteria for dividing triads into classes and species, and provide data on the number of classes and species. The proposed classification opens up new prospects for studying earthquakes.

## 2. Classification

Let us introduce a rectangular matrix $T(M_0)$ for the abbreviated designation of classes and species of main shocks impacts with magnitude $M \geq M_0$:

$$T(M_0) = \begin{pmatrix} \text{TC1} & \text{TC2} \\ \text{TM1} & \text{TM2} \\ \text{TS1} & \text{TS2} \end{pmatrix}. \tag{1}$$

The letter T is an abbreviation for the word 'Triad'. (Recall that the classification of main shocks attacks is identical to the classification of triads.) Similarly, C, M and S are abbreviations for "Classical", "Mirror" and "Symmetrical". The numbers 1 and 2 indicate whether the triad is full or shortened. Based on this feature, all triads are divided into two types: full triads (first column in matrix (1)) and shortened triads (second codon).

Let's choose a basis for classification. Let $N_-$ be the number of foreshocks during a certain time interval before the main shock, and $N_+$ be the number of aftershocks during exactly the same interval after the main shock. The division of triads into classes is based on the relationship between $N_-$ and $N_+$. If $N_- < N_+$, $N_- > N_+$, or $N_- = N_+$, then such a triad, as stated, will be called a classical, mirror, or symmetric triad, respectively. If both numbers $N_-$ and $N_+$ are numbers 4 and 7 are different from zero, then we will call such a triad complete. If at least one of these numbers is equal to zero, then the triad will be called shortened (see Table). The class and specie of the main shock strike is recognized by its belonging to one or another category of triads. Thus we have three classes and six species of main shocks. Each species can be divided into varieties according to magnitude, or the depth of the hypocenter, or other characteristics.

**Table**. Classification criteria

| Classes ||||||
|---|---|---|---|---|---|
| TC || TM || TS ||
| $N_- < N_+$ || $N_- > N_+$ || $N_- = N_+$ ||
| Species ||||||
| TC1 | TC2 | TM1 | TC2 | TS1 | TS2 |
| $N_- \neq 0$ | $N_- = 0$ | $N_+ \neq 0$ | $N_+ = 0$ | $N_\pm \neq 0$ | $N_\pm = 0$ |

The number of triads of a certain type depends on the magnetic field $M_0$ established during the selection of the main shocks. Here are the probabilities of detecting one or another species at $M_0 = 6$ and $M_0 = 7$:

$$T(6) = \begin{pmatrix} 0.063 & 0.392 \\ 0.020 & 0.030 \\ 0.023 & 0.472 \end{pmatrix}, \qquad T(7) = \begin{pmatrix} 0.218 & 0.662 \\ 0.014 & 0.004 \\ 0.005 & 0.097 \end{pmatrix}. \qquad (2)$$

The total number of triads $T(6)$ and $T(7)$ in the period from 1973 to 2019 is 5259 and 565, respectively. The probability values in the $T(6)$ matrix are quite reliable, while in the $T(7)$ matrix the probabilities of occurrence of species TM2 and TS1 are lower than the accuracy of the calculations.

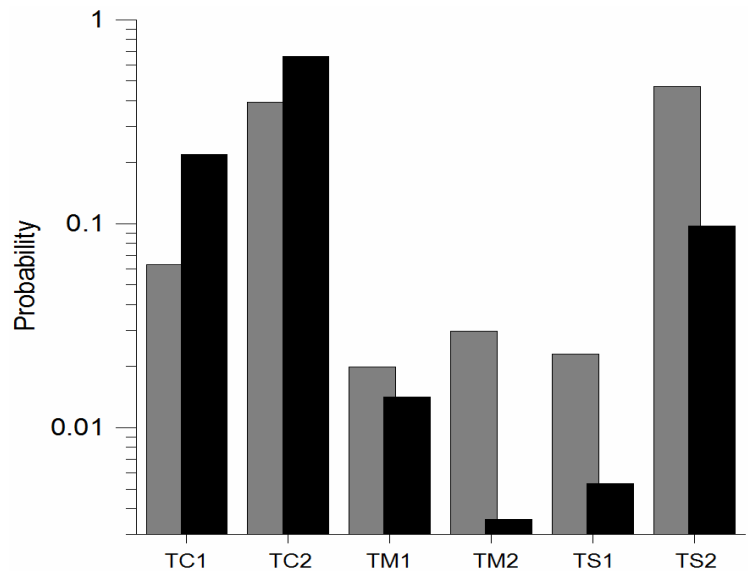

**Fig. 3**. Probability of detecting a triad of a certain type at $M_0 = 6$ (gray) and at $M_0 = 7$ (black).

The information contained in matrices (2) is presented in visual form in Figure 3. We see that the appearance of unusual species TM1, TM2 and TS1 is unlikely, but they really exist and are of some interest. What was unexpected for us was the high probability of the occurrence of main shocks of type TS2. The question of the existence of isolated strong earthquakes, preceded by no foreshocks and not followed by aftershocks, undoubtedly requires additional study.

## 3. Discussion

Our classification of main shocks satisfies the most important classification rule,

according to which the division of a set into subsets should be done only on one basis (attribute). As a basis, we chose the main shock to belong to one of six species of triads. In turn, the classification of triads also satisfies the specified rule, since the division into classes and species is made on the basis of simple arithmetic relationships between the numbers of foreshocks and aftershocks.

Classical triads have always been the object of study in seismology. It has long become commonplace and to some extent understandable that the activity of foreshocks is weaker than the activity of aftershocks. In the review [Guglielmi, 2015], the appearance of foreshocks is compared with the known growth of fluctuations at the threshold of a critical transition of a dynamic system from one state to another. In this light, the discovery of mirror triads forming a small but non-empty subset of the set of all triads is surprising. The complete lack of understanding of the mechanism of occurrence of the mirror triad calls into question our usual judgments about foreshocks and aftershocks in the classical triad. Isn't it an illusion, a common misconception, that we understand at least in the most general terms the origin of foreshocks and aftershocks?

Sometimes the path to truth lies through error, so the question seems quite appropriate to us. The history of science shows that persistent misconceptions often arise in the process of research and in many cases are successfully overcome. We express confidence that a thorough experimental study will make it possible to understand the geophysical meaning of the appearance of mirror analogues of classical triads.

Complete symmetrical TS1 triads are of particular interest due to their beauty and mystery of origin. They occupy a borderline position between the classical and mirror triads. It is curious that at $M \geq 6$, TS1 triads are found only 2.7 times less frequently than full classical TC1 triads (see matrix $T(6)$)

Let's use the information contained in Figure 2 and make a assessment of the activation coefficients $\sigma_-$ and deactivation coefficients $\sigma_+$ of the earthquake source

according to formula

$$\sigma_{\mp} = \mp \frac{d}{dt}\langle g_{\mp}\rangle. \qquad (2)$$

Here $g_{\mp} = 1/n_{\mp}$, $n_{-}$ is the frequency of foreshocks, and $n_{+}$ is the frequency of aftershocks. Corner marks indicate smoothing of functions $g_{\mp} = 1/n_{\mp}$. The $\sigma_{\mp}$ values serve as a qualitative measure of the state of the earthquake source. (For more information about this, see, for example, the reviews [Zavyalov, Zotov, Guglielmi, Klain, 2022; Guglielmi, Klain, Zavyalov, Zotov, 2023].)

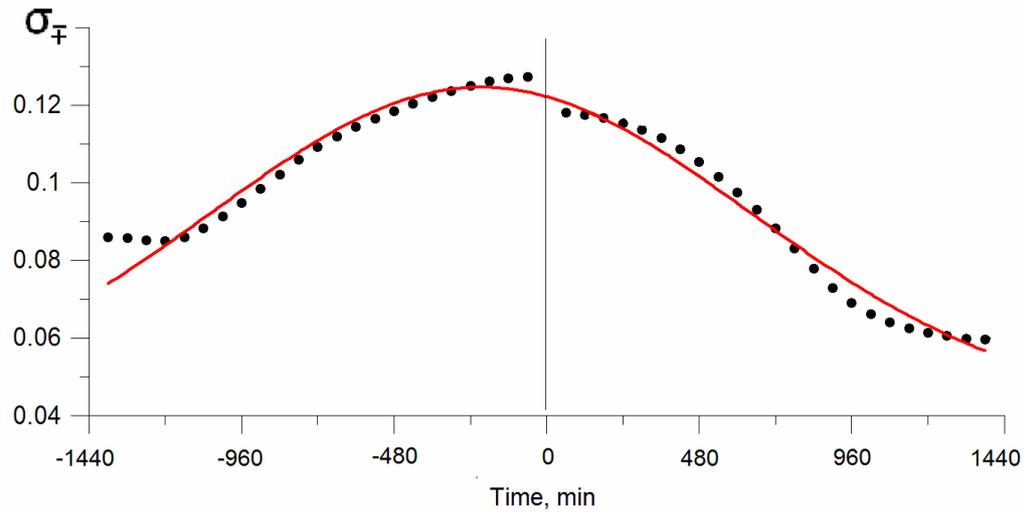

**Fig. 4**. Activation coefficient (to the left of the reference point) and deactivation coefficient of the earthquake source (to the right) during excitation of complete symmetrical triads TS1.

Figure 4 shows the result. The red line shows the approximation of experimental points by Gaussian

$$\sigma_{\mp}(t) = \sigma_0 \exp\left[-t^2/\theta^2\right]. \qquad (3)$$

What is unusual is the monotonous decrease in the deactivation coefficient over time. This distinguishes the symmetrical triad from the classical one, which is characterized by a constant deactivation coefficient.

Main shocks TS2 require separate study. We included TS2 in matrix (1) to ensure that our classification was complete in the sense that the totality of species exhausts the generic concept of the triad. At the same time, we did not exclude the possibility that the set TS2 would be empty. An unexpected discovery was the fact that there is a non-empty set of incomplete symmetric triads.

## 4. Conclusion

We introduced the concept of an earthquake triad as a sequence of foreshocks, mainshocks and aftershocks. There are three classes of triads depending on whether the number of foreshocks is less than, greater than, or equal to the number of aftershocks. We agreed to call the corresponding triads classical, mirror, and symmetrical.

The mainshock is an essential member of the triad, while formshocks and/or aftershocks may not be observed. Therefore, the classification of the main shock coincides with the classification of triads. Each of the three mainshock classes is divided into two species depending on whether the mainshock is part of a full triad containing a non-zero number of foreshocks and aftershocks, or part of a truncated triad in which foreshocks and/or aftershocks are not observed. Thus, each main shock can be classified into one of three classes and one of six species.

In developing the classification, we discovered mirror triads, the existence of which, as far as we know, had not been previously noticed. Mirror triads are few in number, but are of great interest to earthquake physics. It is quite plausible to assume that the conditions for the excitation of main shocks in mirror triads are radically different from the conditions for the excitation of main shocks in classical triads. We believe that the classification of main shocks will open up new perspectives for earthquake research.

*Acknowledgments*. We express our gratitude to B.I. Klain and A.D. Zavyalov for fruitful discussions of the problem. We thank F.Z. Feygin and A.S. Potapov for his interest in the work and support. We thank colleagues at the US Geological Survey for lending us their earthquake catalogs USGS/NEIC for use. The work was carried out according to the plan of state assignments of Schmidt Institute of Physics of the Earth, Russian Academy of Sciences.